\documentclass[prd,twocolumn,showpacs,floatfix,amsmath,nofootinbib,amssymb,floatfix]{revtex4}
\usepackage{graphicx,color,dcolumn,booktabs,bm,multirow}
\usepackage{longtable,lscape}
\usepackage{txfonts}
 \usepackage{enumitem}
\usepackage{overpic}
\usepackage{amssymb}
\usepackage{indentfirst}
\usepackage{feynmf}   
\usepackage{slashed}  
\usepackage{cases}
\usepackage{color}
\usepackage{multirow}
\usepackage{epstopdf}
\usepackage{float}
\usepackage{graphicx,color,dcolumn,booktabs,bm}

\usepackage[colorlinks,
            citecolor=blue,
            anchorcolor=red,
            menucolor=red,
            linkcolor=red,
            filecolor=red,
            runcolor=red,
            urlcolor=blue,
            frenchlinks=red]{hyperref}

\begin{document}

\title{Investigations on the light hadron decays of $Z_b(10610)$ and $Z_b(10650)$}
\author{Qi Wu$^{1}$}\email{wu\_qi@pku.edu.cn}
\author{Yuanxin Zheng$^{2}$}\email{13563834773@163.com}
\author{Shidong Liu$^{2}$ \footnote{Corresponding author}}\email{liusd@qfnu.edu.cn}
\author{Gang Li$^{2}$ \footnote{Corresponding author}}\email{gli@qfnu.edu.cn}
\affiliation{$^1$School of Physics and Center of High Energy Physics, Peking University, Beijing 100871, China\\
$^2$School of Physics and Physical Engineering, Qufu Normal University, Qufu 273165, China}

\begin{abstract}
The light hadron decay processes of $Z_b(10610)/Z_b(10650)$ provide us a way to study their nature and decay
mechanism. In this work, we evaluate the branching ratios of $Z_b(10610)/Z_b(10650) \to VP$ ($V$ and $P$ stand for light vector and pseudoscalar mesons, respectively) using an effective Lagrangian approach, in which the contributions of intermediate bottomed meson triangle loops are considered. In our calculations, the $Z_b(10610)$ and $Z_b(10650)$ are regarded as $B\bar{B}^*+c.c.$ and $B^*\bar{B}^*$ molecular states, respectively. The predicted branching ratios of $Z_b(10610)\rightarrow VP$ are about in the order of $10^{-2}$, while the branching ratios of $Z_b(10650)\rightarrow VP$ are in the order of $10^{-3}$. Furthermore, the dependence of these ratios between different decay modes of $Z_b(10610)/Z_b(10650)$ on the mixing $\eta-\eta^\prime$ angle $\theta_P$ is investigated, which may be a good quantity for the experiments. It is hoped that the calculations here could be tested by future experiments.
\end{abstract}

\date{\today}
\maketitle

\section{Introduction}
\label{sec:introduction}

The discovery of $X(3872)$ in 2003 opens the gate to the $XYZ$ states in the heavy quarkonium region~\cite{Belle:2003nnu}. A large amount of experimental and theoretical studies have been devoted to the $XYZ$ states (see Refs.~\cite{Chen:2016qju,Hosaka:2016pey,Lebed:2016hpi,Esposito:2016noz,Guo:2017jvc,Ali:2017jda,Olsen:2017bmm,Karliner:2017qhf,Yuan:2018inv,Dong:2017gaw,Liu:2019zoy, Chen:2022asf,Meng:2022ozq} for recent reviews). Many of them cannot be accommodated in the conventional quark model as $Q\bar Q$ and thus turn out to be excellent candidates for exotic states.  

In the charm sector, many charmoniumlike states, such as $Z_c
(3900)/Z_c(4020)/Z_{cs}(3985)$~\cite{BESIII:2013ris,BESIII:2020qkh}, $T^+_{cc}(3875)$~\cite{LHCb:2021vvq,LHCb:2021auc}, $X(6900)$~\cite{LHCb:2020bwg}, $P_c/P_{cs}$~\cite{LHCb:2015yax,LHCb:2019kea,LHCb:2021chn,LHCb:2020jpq,Collaboration:2022boa}, have been observed experimentally. However, in the bottom sector, only two bottomonium-like states $Z_b(10610)$ and $Z_b(10650)$, hereafter referred to $Z_b^{\pm, 0}$ and $Z^{\prime \pm, 0}_b$ respectively, were observed in the invariant mass distributions of the $\pi^\pm \Upsilon(nS)$ $(n=1,2,3)$ and $\pi^\pm h_b(mP)$ $(m=1,2)$ final states in the processes $\Upsilon(5S)\rightarrow\pi^\pm \pi^\mp \Upsilon(nS)$ and $\Upsilon(5S)\rightarrow\pi^\pm \pi^\mp h_b(mP)$~\cite{Belle:2011aa} by the Belle Collaboration in 2011. The charged pion angular distribution analysis~\cite{Adachi:2011mks} indicates that the quantum number of the two states favors $I^G(J^P)=1^+(1^+)$. Furthermore, the amplitude analysis of $e^+ e^-\rightarrow\Upsilon(nS)\pi^+ \pi^-$~\cite{Belle:2014vzn} confirms $I^G(J^P)=1^+(1^+)$ for $Z_b$ and $Z^\prime_b$. The neutral state $Z^0_b$ was observed in a Dalitz analysis of $\Upsilon(5S)\rightarrow\pi^0 \pi^0 \Upsilon(nS)$ decays by the Belle Collaboration~\cite{Belle:2012glq,Belle:2013urd}, which indicates that $Z^{{(\prime)}}_b$ are isovector states. Therefore, they contain at least four constituent quarks and thus are ideal candidates of the exotic hadronic state.

The peculiar natures of $Z^{{(\prime)}}_b$ intrigued theorists to explore their inner structure. Since the masses of $Z_b$ and $Z^\prime_b$ are very close to the thresholds of $B\bar{B}^*+c.c.$ and $B^*\bar{B}^*$ respectively, they are naturally regarded as the deuteronlike molecular states composed of $B\bar{B}^*+c.c.$ and $B^*\bar{B}^*$~\cite{Bondar:2011ev,Sun:2011uh,Mehen:2011yh,Cleven:2011gp,Dias:2014pva,Li:2012wf,Yang:2011rp,Zhang:2011jja,Wang:2013daa, Wang:2014gwa,Dong:2012hc,Ohkoda:2013cea,Li:2012as,Cleven:2013sq,Li:2012uc,Li:2014pfa,Xiao:2017uve,Wu:2018xaa,Wu:2020edh,Voloshin:2011qa}, which could explain most of the properties of $Z_b^{(\prime)}$. The authors in Ref.~\cite{Bondar:2011ev} pointed out that the observations with similar rates of the $Z_b^{(\prime)}$ in both the final states containing the spin-triplet $\Upsilon(1S,2S,3S)$ and spin-singlet $h_b(1P,2P)$ can be naturally understood since the $b\bar{b}$ pairs in both $Z_b$ and $Z_b^\prime$ are mixtures of a spin-triplet and a spin-singlet in the molecular states scenario. In the framework of the one-boson-exchange model, $Z_b$ and $Z^\prime_b$ 
can be interpreted as the $B\bar{B}^*$ and $B^*\bar{B}^*$ molecular states~\cite{Sun:2011uh}. In the molecular states picture, the masses of $Z^{{(\prime)}}_b$ could be well reproduced using QCD sum rules~\cite{Zhang:2011jja,Wang:2013daa,Wang:2014gwa}. Besides, many studies of the  decays~\cite{Dong:2012hc,Ohkoda:2013cea,Li:2012as,Cleven:2013sq,Li:2012uc,Li:2014pfa,Xiao:2017uve,Wu:2020edh} and productions~\cite{Wu:2018xaa,Voloshin:2011qa} of $Z_b^{(\prime)}$ also support the molecule interpretation. It seems that $Z^{(\prime)}_b$ are molecular states composed of $B\bar{B}^*+c.c.$ and $B^*\bar{B}^*$, but other interpretations could not be ruled out. For example, it is interpreted as the tetraquark states with four valence quarks ($b\bar{b}q\bar{q}, q=u,d$) since $Z^{(\prime)}_b$ were observed in the $\pi \Upsilon(1S,2S,3S)$ and $\pi h_b(1P,2P)$ invariant mass spectrum. 
With the help of various models, different possible tetraquark state configurations of $Z^{(\prime)}_b$ were investigated~\cite{Guo:2011gu,Cui:2011fj,Wang:2013zra}. Besides the QCD exotic interpretations, the structures corresponding to $Z^{(\prime)}_b$ could be reproduced through initial-single-pion-emission mechanism~\cite{Chen:2011pv,Chen:2012yr} or cusp effect~\cite{Bugg:2011jr}.

In addition to the mass spectrum studies, the decays and productions also contain detailed dynamical information and hence provide another perspective about their properties. The productions of $Z^{{(\prime)}}_b$ states from the $\Upsilon(5S)$ radiative decays and hidden bottom decays were studied with $Z_b$ and $Z^{\prime}_b$ being $B\bar{B}^*+c.c.$ and $B^*\bar{B}^*$ hadronic molecules, respectively ~\cite{Voloshin:2011qa,Wu:2018xaa}.  The hidden-bottom and radiative decays of $Z^{{(\prime)}}_b$ have been extensively investigated  using various methods. $Z^{{(\prime)}}_b \to \Upsilon(nS)\pi$ was evaluated in a phenomenological Lagrangian approach in Ref.~\cite{Dong:2012hc}. QCD multipole expansion was applied to study the ratios of the decay rates of $Z^{{(\prime)}}_b \to \Upsilon(nS)\pi$ and $Z^{{(\prime)}}_b \to h_b(mP)\pi$ in Ref.~\cite{Li:2012uc}. By adopting an effective Lagrangian approach, the authors in Refs.~\cite{Li:2012as,Ohkoda:2013cea,Xiao:2017uve} evaluated the bottom meson loop contributions of $Z^{{(\prime)}}_b \to \Upsilon(nS)\pi$, $Z^{{(\prime)}}_b \to h_b(mP)\pi$, and $Z^{(\prime)}_b\rightarrow\eta_b(mP)\gamma$. The decays $Z^{{(\prime)}}_b \to \Upsilon(nS)\pi$, $Z^{{(\prime)}}_b \to h_b(mP)\pi$, and $Z^{(\prime)}_b\rightarrow \chi_{bJ}(mP)\gamma$ were investigated within a nonrelativistic effective field theory in Ref.~\cite{Cleven:2013sq}.

\begin{table}[tb]
\centering
\renewcommand{\arraystretch}{1.2}
    \caption{Measurements of the branching ratios ($\%$) of $Z_b$ and $Z^\prime_b$ from PDG~\cite{Workman:2022ynf}.}\label{tab:br}
    \setlength{\tabcolsep}{6.mm}
\begin{tabular}{ccc}
  \toprule[1pt]\toprule[1pt]
  Decay channels & $Z_b^+$ & $Z_b^{\prime +}$ \\
  \midrule[1pt]
  $\Upsilon(1S)\pi^+$ & $0.54^{+0.19}_{-0.15}$ & $0.17^{+0.08}_{-0.06}$ \\
  $\Upsilon(2S)\pi^+$ & $3.6^{+1.1}_{-0.8}$ & $1.4^{+0.6}_{-0.4}$ \\
  $\Upsilon(3S)\pi^+$ & $2.1^{+0.8}_{-0.6}$ & $1.6^{+0.7}_{-0.5}$ \\
  $h_b(1P)\pi^+$ & $3.5^{+1.2}_{-0.9}$ & $8.4^{+2.9}_{-2.4}$ \\
  $h_b(2P)\pi^+$ & $4.7^{+1.7}_{-1.3}$ & $15\pm4$ \\
  $B^+\bar{B}^{*0}+B^{*+}\bar{B}^0$ & $85.6^{+2.1}_{-2.9}$ & ... \\
  $B^{*+}\bar{B}^{*0}$& ... & $74^{+4}_{-6}$ \\
  \bottomrule[1pt]\bottomrule[1pt]
\end{tabular}
\end{table}

Besides the resonance parameters of the $Z^{{(\prime)}}_b$ states, the Belle Collaboration also measured the open-bottom and hidden-bottom decay modes of $Z^{\pm}_b$ and $Z^{\pm}_b$ in $\Upsilon(5S)$ decays~\cite{Belle:2012koo}. In Table~\ref{tab:br}, we list the branching ratios of $Z_b^+$ and $Z^{\prime +}_b$ obtained from Particle Data Group (PDG)~\cite{Workman:2022ynf}. It can be seen that the dominant decay modes of $Z_b^+/Z^{\prime +}_b$ are open-bottom decays, i.e., $Z_b^+$ and $Z^{\prime +}_b$ mainly decay into $B^+\bar{B}^{*0} + c.c.$ with branching ratio $(85.6^{+2.1}_{-2.9})\%$ and $B^{*+}\bar{B}^{*0}$ with branching ratio $74^{+4}_{-6}\%$, respectively. The secondary decay modes of $Z_b^+$ and $Z^{\prime +}_b$ are the hidden-bottom decay, namely $\pi^+ \Upsilon(nS)(n=1,2,3)$ and $\pi^+ h_b(mP)(m=1,2)$. The branching ratios are $14.4^{+2.5}_{-1.9}\%$ for $Z^+_b\rightarrow (b\bar{b})+\pi^+$ and $26.6^{+5.0}_{-4.7}\%$ for $Z^{\prime+}_b\rightarrow (b\bar{b})+\pi^+$. To better understand the nature of $Z_b^{(\prime)}$, the study of other decay modes is necessary. For example, the light hadron decay modes can provide a good platform to study their nature. In Refs.~\cite{Wu:2016ypc,Wang:2022qxe}, we have investigated the charmless decays of charmoniunlike states $Z_c(3900)/Z_c(4020)$ and $X(3872)$ by considering the contributions of intermediate charmed meson loops, and predicted accessible decay rates. In this work, we investigate the light hadron decay modes of $Z^{(\prime)}_b\to VP$ via the intermediate bottomed meson loops using an effective Lagrangian approach. We will focus on the light hadron decays of $Z_b^{+}$ and $Z^{\prime+}_b$. For simplicity, we do not distinguish the charged and the neutral $Z^{(\prime)}_b$ states and use $Z^{(\prime)}_b$ to represent $Z^{(\prime)+}_b$.

This article is organized as follows. After the Introduction, we present the theoretical framework used in Sec.~\ref{sec:Sec2}. The numerical results and discussion are presented in Sec.~\ref{Sec:Num}, and a brief summary is given in Sec.~\ref{sec:summary}.

\section{Theoretical framework}
\label{sec:Sec2}

\begin{figure}[htb!]
\begin{tabular}{cc}
  \centering
 \includegraphics[width=3.2cm]{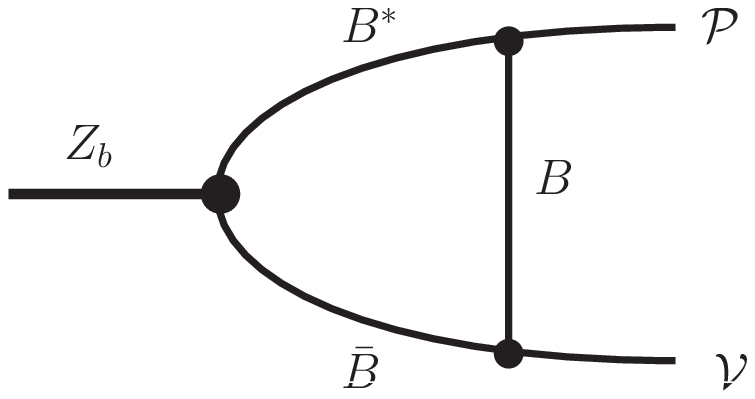} &  \includegraphics[width=3.2cm]{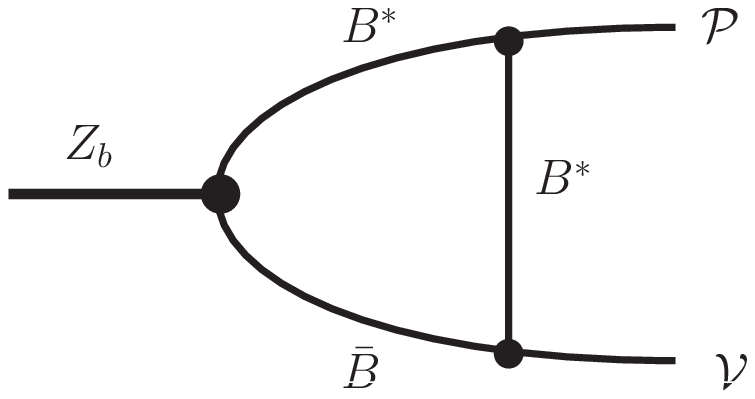} \\
 $(a)$ & $(b)$ \\ \\
 \includegraphics[width=3.2cm]{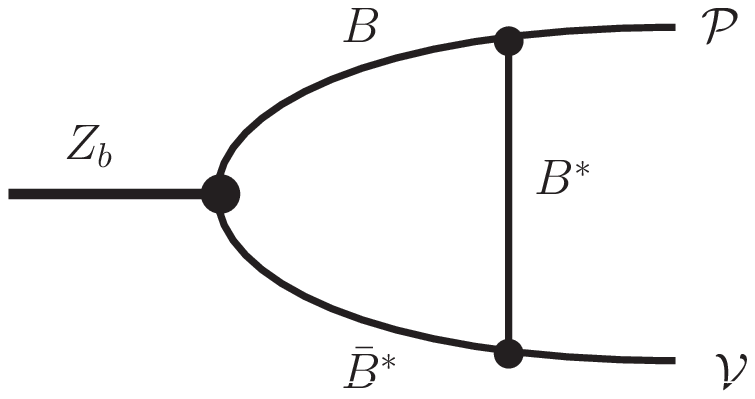} &\\
   $(c)$ &\\
 \end{tabular}
  \caption{The hadron-level diagrams contributing to the light hadron decays $Z_{b} \rightarrow VP$. The charge conjugated diagrams are
not shown but included in our calculations.}\label{Fig:Tri1}
\end{figure}

\begin{figure}[htb!]
\begin{tabular}{ccc}
  \centering
 \includegraphics[width=3.2cm]{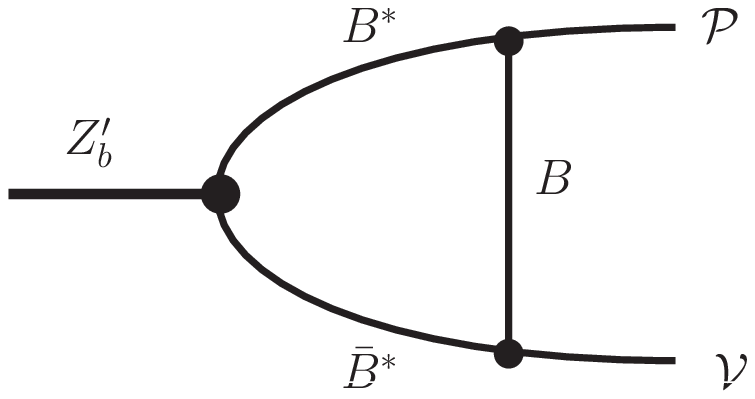}&
 \includegraphics[width=3.2cm]{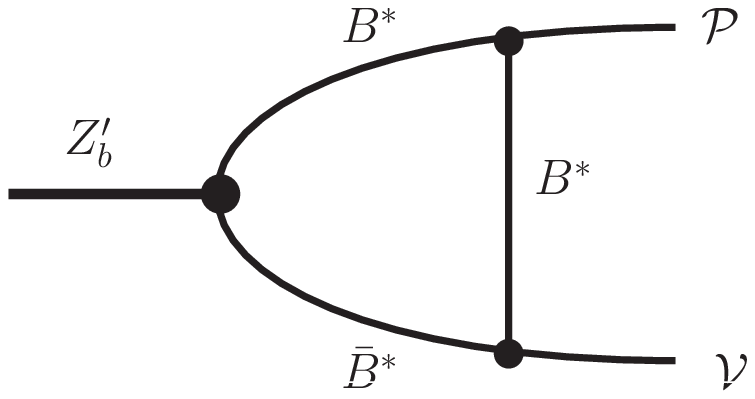}\\
 $(a)$ & $(b)$ \\
 \end{tabular}
  \caption{The hadron-level diagrams contributing to the light hadron decays $Z^\prime_b \rightarrow VP$. The charge conjugated diagrams are
not shown but included in our calculations.}\label{Fig:Tri2}
\end{figure}

We study the light hadron decays of $Z^{(\prime)}_b$ states using the effective Lagrangian approach. The experimental measurements reveal that $Z_b$ and $Z^\prime_b$ dominantly decay into $B\bar{B}^*+c.c.$ and $B^*\bar{B}^*$, leading to strong couplings of $Z_b$ and $Z^\prime_b$ to $B\bar{B}^*+c.c.$ and $B^*\bar{B}^*$. Their light hadron decays can be proceeded by the triangle diagrams as shown in Figs.~\ref{Fig:Tri1} and \ref{Fig:Tri2}, each of which has three bottomed mesons in the triangle loop. In the following, we apply such a mechanism to study the light hadron decays of $Z_b$ and $Z^\prime_b$.

\subsection{Effective Lagrangian}

In order to calculate the triangle loops shown in Figs. \ref{Fig:Tri1} and \ref{Fig:Tri2}, we need the effective couplings of the $Z_b$ and $Z^\prime_b$ states to $B {\bar B}^*$ and $B^* {\bar B}^*$ channels, respectively. The effective Lagrangians describing the couplings of
$Z_b$ and $Z^\prime_b$ states to $B {\bar B}^*$ and $B^* {\bar B}^*$ channels via $S$ wave are written as~\cite{Cleven:2013sq}
\begin{eqnarray}
\mathcal{L}_{Z_b^{(\prime)} B^{(*)} B^{(*)}}&=&g_{Z_bBB^*}Z^\mu_b (BB^{*\dag}_\mu +B^*_\mu B^\dag) \nonumber \\
&& + g_{Z^\prime_bB^*B^*}\epsilon_{\mu\nu\alpha\beta}\partial^\mu Z^{\prime\nu}_b B^{*\alpha} B^{*\dag\beta}+H.c. \, ,\label{Eq:zb}
\end{eqnarray}
where ${B}^{(\ast)}=(B^{(\ast)+}, B^{(\ast)0},B^{(\ast)0}_s)$ is the bottom
meson triplets; $g_{Z_b BB^*}$ and $g_{Z^\prime_b B^*B^*}$ are the couplings constants. We use the convention $\varepsilon_{0123}=+1$. With the above effective Lagrangians, we can obtain 
\begin{eqnarray}
\Gamma_{Z_b\to B^+{\bar B}^{*0} + B^{*+} {\bar B}^0} &=& \frac{|\vec{q}_B|}{12\pi m^2_{Z_b}} g^2_{Z_b BB^*}\Big(\frac{|\vec{q}_B|^2}{m^2_{B^*}}+3\Big)\, , \\
\Gamma_{Z^\prime_b\to B^{*+}{\bar B}^{*0} } &=& \frac{|\vec{q}_{B^*}|}{24\pi }g^2_{Z^\prime_b B^* B^*}\Big(\frac{m^2_{Z^\prime_b}}{m^2_{B^*}}+2\Big)\, ,
\end{eqnarray}
with ${\vec q}_B$ and ${\vec q}_{B^*}$ being the three-momenta of the $B$ and $B^*$ mesons in the rest frame of $Z_b$ and $Z^\prime_b$, respectively. With the experimentally measured branching ratios $BR({Z_b\to B^+{\bar B}^{*0} + B^{*+} {\bar B}^0})=85.6\% $, $BR(Z^\prime_b\to B^{*+}{\bar B}^{*0}) =74\% $ shown in Table~\ref{tab:br}, and the center values of the total widths of $Z^{(\prime)}_b$, we can get the relevant couplings $g_{Z_b BB^*} =13.23$~GeV and $g_{Z^\prime_b B^*B^*}=0.94 $, respectively. 

On the other hand, the Lagrangians relevant to the light vector and pseudoscalar mesons can be constructed based on the heavy quark limit and chiral symmetry~\cite{Casalbuoni:1996pg,Colangelo:2003sa,Cheng:2004ru}
\begin{eqnarray}
 {\cal L} &=& -ig_{B^{\ast }B
{\mathcal P}}\left( B^{\dag}_i \partial_\mu {\mathcal P}_{ij} B_j^{\ast \mu}-B_i^{\ast \mu\dagger} \partial_\mu {\mathcal P}_{ij}  B_j\right) \nonumber \\ 
&&+\frac{1}{2}g_{B^\ast B^\ast {\mathcal P}}\varepsilon _{\mu
\nu \alpha \beta }B_i^{\ast \mu \dag}\partial^\nu {\mathcal P}_{ij}  {\overset{
\leftrightarrow }{\partial }}{\!^{\alpha }} B_j^{\ast \beta } - ig_{{B}{B}\mathcal{V}} {B}_i^\dagger {\stackrel{\leftrightarrow}{\partial}}{\!_\mu} {B}^j(\mathcal{V}^\mu)^i_j \nonumber \\
&& -2f_{{B}^*{B}\mathcal{V}} \epsilon_{\mu\nu\alpha\beta}
(\partial^\mu \mathcal{V}^\nu)^i_j
({B}_i^\dagger{\stackrel{\leftrightarrow}{\partial}}{\!^\alpha} {B}^{*\beta j}-{B}_i^{*\beta\dagger}{\stackrel{\leftrightarrow}{\partial}}{\!^\alpha} {B}^j) \nonumber \\ && + ig_{{B}^*{B}^*\mathcal{V}} {B}^{*\nu\dagger}_i {\stackrel{\leftrightarrow}{\partial}}{\!_\mu} {B}^{*j}_\nu(\mathcal{V}^\mu)^i_j \nonumber \\ &&+4if_{{B}^*{B}^*\mathcal{V}} {B}^{*\dagger}_{i\mu}(\partial^\mu \mathcal{V}^\nu-\partial^\nu
\mathcal{V}^\mu)^i_j {B}^{*j}_\nu +{\rm H.c.} \,. \label{Eq:light-meson}
\end{eqnarray}
Again, ${B}^{(\ast)}=(B^{(\ast)+}, B^{(\ast)0},B^{(\ast)0}_s)$ is the bottom
meson triplets; $\mathcal P$ and ${\mathcal V}_\mu$ are $3\times 3$ matrices representing the pseudoscalar and vector mesons, and their specific forms are
\begin{widetext}
\begin{eqnarray}
    \label{eq-13}
    \mathcal{P} =
\left(\begin{array}{ccc}
\frac{\pi^{0}}{\sqrt 2}+ \frac{\eta\cos{\alpha_P}+\eta^\prime\sin{\alpha_P}}{\sqrt{2}} &\pi^{+} &K^{+}\\
\pi^{-} &-\frac{\pi^{0}}{\sqrt 2}+ \frac{\eta\cos{\alpha_P}+\eta^\prime\sin{\alpha_P}}{\sqrt{2}} &K^{0}\\
K^{-} &\bar K^{0} &-\eta\sin{\alpha_P}+\eta^\prime\cos{\alpha_P}\end{array}\right),
\mathcal{V} = \left(\begin{array}{ccc}\frac{\rho^0} {\sqrt {2}}+\frac {\omega} {\sqrt {2}}&\rho^+ & K^{*+} \\
\rho^- & -\frac {\rho^0} {\sqrt {2}} + \frac {\omega} {\sqrt {2}} & K^{*0} \\
K^{*-}& {\bar K}^{*0} & \phi \\
\end{array}\right) \, .
\end{eqnarray}
\end{widetext}
The physical states $\eta$ and $\eta^\prime$ are the mixing of favor eigenstates $n\bar{n}=(u\bar{u}+d\bar{d})/\sqrt{2}$ and $s\bar{s}$, which have the following wave functions,
\begin{eqnarray}
|\eta\rangle &=&\cos{\alpha_P}|n\bar{n}\rangle-\sin{\alpha_P}|s\bar{s}\rangle,\nonumber \\
|\eta^\prime\rangle &=&\sin{\alpha_P}|n\bar{n}\rangle+\cos{\alpha_P}|s\bar{s}\rangle \, , \label{eq:eta-etap mixing}
\end{eqnarray}
where $\alpha_P\simeq \theta_P+\arctan\sqrt{2}$. The empirical value for the
pseudoscalar mixing angle $\theta_P$ is in the range $[-24.6^\circ , -11.5^\circ]$~\cite{Workman:2022ynf}. 


In the heavy quark and chiral limits, the couplings of bottomed mesons to the light vector mesons have the following relationships~\cite{Casalbuoni:1996pg,Cheng:2004ru},
\begin{eqnarray}
g_{{ B}{ B}V} = g_{{ B}^*{ B}^*V}=\frac{\beta g_V}{\sqrt{2}} , \quad f_{{ B}^*{ B}V}=\frac{ f_{{ B}^*{ B}^*V}}{m_{{ B}^*}}=\frac{\lambda g_V}{\sqrt{2}} \, , \\
g_{{B}^{*} {B} \mathcal{P}}=\frac{2 g}{f_{\pi}} \sqrt{m_{{B}} m_{{B}^{*}}}, \quad g_{{B}^{*} {B}^{*} {P}}=\frac{g_{{ B}^{*} { B} {\mathcal {P}}}}{\sqrt{m_{{B}} m_{{B}^{*}}}} \, .
\end{eqnarray}
In this work, we take the parameters $\beta=0.9$, $\lambda = 0.56 \, {\rm GeV}^{-1} $, $g=0.59$, and $g_V = {m_\rho /f_\pi}$ with $f_\pi = 132$ MeV as used in previous works~\cite{Casalbuoni:1996pg,Isola:2003fh}.

\subsection{Decay Amplitude}

With the above effective Lagrangians, the decay amplitudes of these triangle diagrams in Figs.~\ref{Fig:Tri1} and \ref{Fig:Tri2} can be easily obtained. For $Z_b (p_1)  \to [B^{(*)} (q_1) {\bar B}^{(*)} (q_3)] B^{(*)} (q_2) \rightarrow V (p_2) P(p_3)$  shown in Fig.~\ref{Fig:Tri1}, the explicit amplitudes are
\begin{eqnarray}
\mathcal{M}_{a}^{Z_b}&=&i^3 \int\frac{d^4 q}{(2\pi)^4}\Big[g_{Z_bBB^*} \epsilon_{1\mu}\Big]\Big[g_{B^\ast BP}p_{3\nu}\Big]
\Big[g_{BBV}\epsilon_{2\alpha}^* (q^\alpha_3-q_2^\alpha)\Big]\nonumber\\
&&\times\frac{-g^{\mu\nu}+q^\mu_1 q^\nu_1 /m^2_1}{q^2_1-m^2_1}\frac{1}{q^2_2-m^2_2}\frac{1}{q_3^2-m_3^2}\prod_i \mathcal{F}_i \left(q_i^{2}\right)\nonumber\,,\\
\mathcal{M}_{b}^{Z_b}&=&i^3 \int\frac{d^4 q}{(2\pi)^4}\Big[g_{Z_bBB^*} \epsilon_{1\kappa}\Big]\Big[\frac{1}{2}g_{B^\ast B^\ast P}\varepsilon_{\mu\nu\alpha\beta}p^\nu_3 (q^\alpha_1+q_2^\alpha)\Big]\nonumber\\
&&\times\Big[-2f_{B^\ast BV}\varepsilon_{\rho\sigma\tau\xi}p^\rho_2 \epsilon^{*\sigma}_2 (q_2^\tau-q^\tau_3)\Big]\nonumber\\
&&\times\frac{-g^{\kappa\beta}+q^\kappa_1 q^\beta_1 /m^2_1}{q^2_1-m^2_1}\frac{-g^{\mu\xi}+q_2^{\mu}q_2^{\xi}/m^2_2}{q_2^2-m^2_2} \frac{1}{q^2_3-m^2_3}\prod_i \mathcal{F}_i \left(q_i^{2}\right)\nonumber\,,\\
\mathcal{M}_{c}^{Z_b}&=&i^3 \int\frac{d^4 q}{(2\pi)^4}\Big[g_{Z_bBB^*} \epsilon_{1\rho} \Big]\Big[-g_{B^\ast BP}p_{3\sigma}\Big]\nonumber\\
&&\times\Big[g_{B^\ast B^\ast V}g_{\tau\theta} (q_{2\kappa}-q_{3\kappa})\epsilon^{*\kappa}_2
+4if_{B^\ast B^\ast V}(-p_{2\tau} \epsilon_{2\theta}^*+p_{2\theta} \epsilon_{2\tau}^*)\Big]\nonumber\\
&&\times \frac{1}{q^2_1-m^2_1}\frac{-g^{\sigma\theta}+q_2^{\sigma} q_2^{\theta} /m^2_2}{q_2^2-m^2_2}\frac{-g^{\rho\tau}+q^{\rho}_3 q^{\tau}_3 /m^2_3}{q^2_3-m^2_3}\prod_i\mathcal{F}_i \left(q_i^{2}\right).\nonumber\\
\end{eqnarray}

The explicit transition amplitudes for $Z^\prime_b (p_1)  \to [B^{*} (q_1) {\bar B}^{*} (q_3)] B^{(*)} (q_2) \rightarrow V (p_2) P(p_3)$   in Fig.~\ref{Fig:Tri2}  are
\begin{eqnarray}
\mathcal{M}_{a}^{Z_b^\prime}&=&i^3 \int\frac{d^4 q}{(2\pi)^4}\Big[ig_{Z^\prime_b B^*B^*} \varepsilon_{\mu\nu\alpha\beta}p_1^\mu \epsilon_1^\nu \Big]\Big[g_{B^\ast BP}p_{3\lambda}\Big]\nonumber\\
&&\times\Big[2f_{B^\ast BV}\varepsilon_{\rho\sigma\tau\xi}p^\rho_2 \epsilon^{*\sigma}_2 (q_2^\tau-q^\tau_3)\Big]\nonumber\\
&&\times\frac{-g^{\alpha\lambda}+q^\alpha_1 q^\lambda_1 /m^2_1}{q^2_1-m^2_1}\frac{1}{q_2^2-m^2_2}\frac{-g^{\beta\xi}+q^\beta_3 q^\xi_3 /m^2_3}{q^2_3-m^2_3}\prod_i \mathcal{F}_i \left(q_i^{2}\right) \, ,\nonumber
\end{eqnarray}
\begin{eqnarray}
\mathcal{M}_{b}^{Z_b^\prime}&=&i^3 \int\frac{d^4 q}{(2\pi)^4}\Big[ig_{Z^\prime_b B^*B^*} \varepsilon_{\mu\nu\alpha\beta}p_1^\mu \epsilon_1^\nu \Big]\Big[\frac{1}{2}g_{B^\ast B^\ast P}\varepsilon_{\rho\sigma\lambda\xi}p^\sigma_3 (q^\lambda_1+q_2^\lambda)\Big]\nonumber\\
&&\times\Big[g_{B^\ast B^\ast V}g_{\tau\theta} (q_{2\kappa}-q_{3\kappa})\epsilon^{*\kappa}_2+4f_{B^\ast B^\ast V}(p_2{\theta} \epsilon_{2\tau}^*-p_{2\tau} \epsilon_{2\theta}^*)\Big]\nonumber\\
&&\times \frac{-g^{\alpha\xi}+q^\alpha_1 q^\xi_1 /m^2_1}{q^2_1-m^2_1}
\frac{-g^{\rho\theta}+q_2^{\rho} q_2^{\theta} /m^2_2}{q_2^2-m^2_2} \frac{-g^{\beta\tau}+q^{\beta}_3 q^{\tau}_3 /m^2_3}{q^2_3-m^2_3} \nonumber\\ && \times \prod_i\mathcal{F}_i \left(q_i^{2}\right)\, . 
\end{eqnarray}
Here $p_1$ ($\varepsilon_1$), $p_2$ ($\varepsilon_2^*$), and $p_3$
 are the four-momenta (polarization vector) of the
initial state $Z^{(\prime)}_b$, final state vector, and pseudoscalar mesons,
respectively. $q_1$, $q_2$, and $q_3$ are the four-momenta of the up, right, and down bottomed mesons in the triangle loop, respectively.

In the present work, we adopt the product of monopole form factors for each intermediate meson, which is
\begin{eqnarray}
\prod_i \mathcal{F}_i \left(q_i^{2}\right) = \mathcal{F}_1 \left(q^2_1\right)\mathcal{F}_2 \left(q_2^2\right)\mathcal{F}_3 \left(q_3^2\right)\label{Eq:FFs}
\end{eqnarray}
with
\begin{eqnarray}
\mathcal{F}_i\left(q_i^{2}\right)=\frac{m^{2}-\Lambda^{2}}{q^{2}_i-\Lambda^{2}} \, .
\end{eqnarray}
The parameter $\Lambda$ can be further reparameterized as $\Lambda_{B^{(\ast)}}=m_{B^{(\ast)}}+\alpha\Lambda_{\rm QCD} $ with $\Lambda_{\rm QCD}=0.22 \ {\rm GeV}$ and $m_{B^{(\ast)}}$ is the mass of the intermediate bottomed meson. The dimensionless model parameter $\alpha$ is of order of unity~\cite{Tornqvist:1993vu,Tornqvist:1993ng,Locher:1993cc,Li:1996yn} but its exact value cannot be obtained form the first principle. In practice, the value of $\alpha$ is usually determined by comparing the theoretical calculations with the corresponding experimental measurements.

\section{Numerical Results and discussions}
\label{Sec:Num}

\begin{figure}[b]
  \centering
 \includegraphics[width=6.5cm]{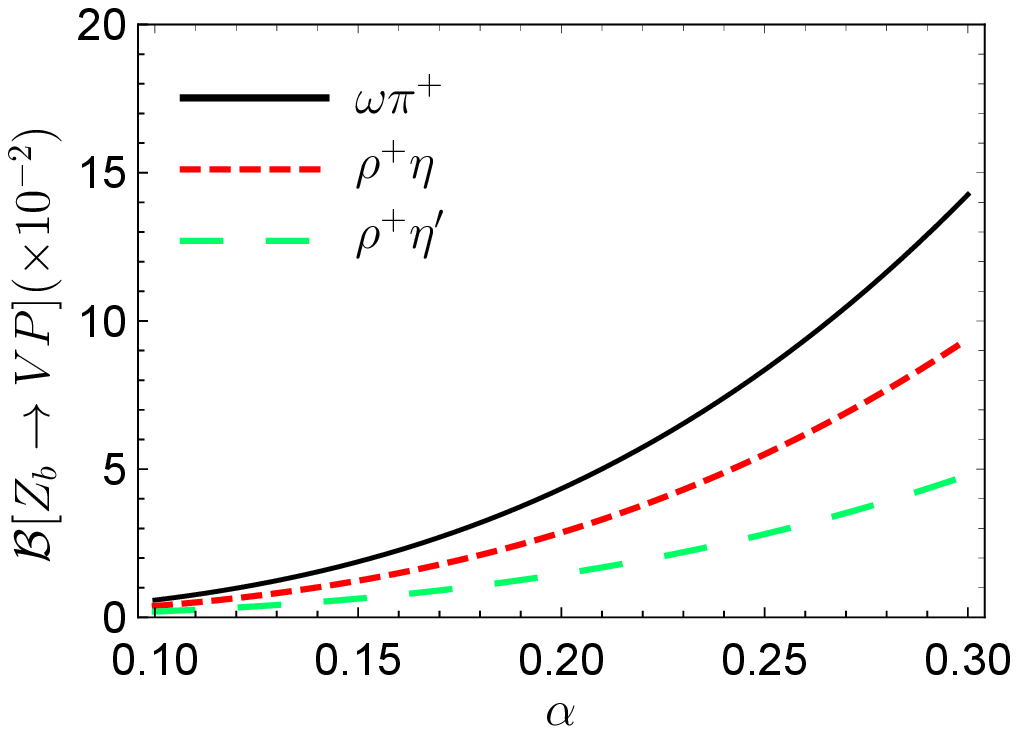}
  \caption{The $\alpha$ dependence of the branching ratios of $Z_b\rightarrow VP$. Here the $\eta-\eta^\prime$
mixing angle $\theta_P=-19.1^\circ$ is taken from Refs.~\cite{MARK-III:1988crp,DM2:1988bfq}. }\label{Fig:B1}
\end{figure}

\begin{figure}[htb]
  \centering
 \includegraphics[width=6.5cm]{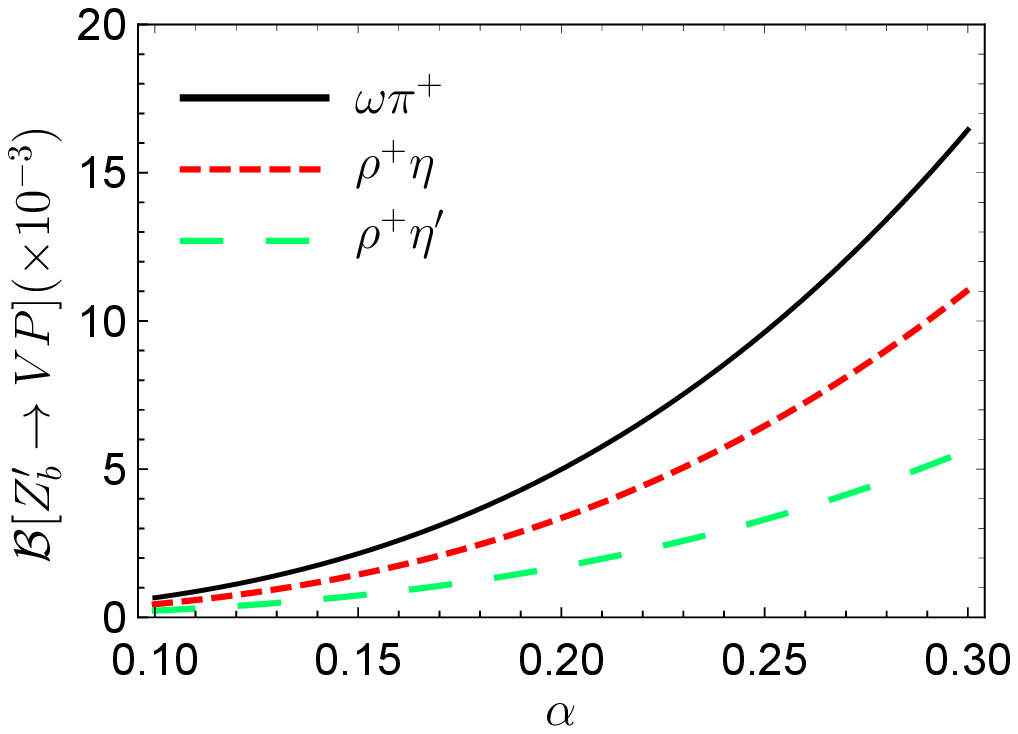}
  \caption{The $\alpha$ dependence of the branching ratios of $Z_b^{\prime} \rightarrow VP$. Here the $\eta-\eta^\prime$
mixing angle $\theta_P=-19.1^\circ$ is taken from Refs.~\cite{MARK-III:1988crp,DM2:1988bfq}.}\label{Fig:B2}
\end{figure}

In this section, we present our numerical results for $Z^{(\prime)}_b \to VP$. The only model parameter in our model is the cutoff parameter $\alpha$ introduced in Eq.~(\ref{Eq:FFs}). No light hadron decay modes of $Z_b$ and $Z^{\prime}_b$ are known so far. For the processes $Z^{(\prime)}_b \to VP$, we expect a relatively small value based on the following reasons. On the one hand, since the masses of $Z_b$ and $Z_b^\prime$ are very close to the thresholds of $B {\bar B}^*$ and $B^* {\bar B}^*$, respectively, the threshold effects would be significantly enhanced. On the other hand, the phase space of the light hadron decay of $Z^{(\prime)}_b$ states is very large. We therefore take $\alpha=0.1\sim0.3$ to estimate the light hadron decays of $Z^{(\prime)}_b$. 

In Figs.~\ref{Fig:B1} and \ref{Fig:B2}, we present the $\alpha$ dependence of the branching ratios of $Z_b\rightarrow VP$ and $Z_b^{\prime}\rightarrow VP$ with the $\eta-\eta^\prime$ mixing angle $\theta_P=-19.1^\circ$, respectively. In the range of $\alpha=0.1 \sim 0.3$, the predicted branching ratios of $Z_b\rightarrow VP$ are about in the order of $10^{-2}$, while for $Z^{\prime}_b\rightarrow VP$ they are in the order of $10^{-3}$. The predicted branching ratios of $Z_b\rightarrow VP$ are about 1 order of magnitude larger than those of $Z^\prime_b\rightarrow VP$. As shown in Figs.~\ref{Fig:Tri1} and \ref{Fig:Tri2}, there are three kinds of diagrams for $Z_b\rightarrow VP$, while there are only two kinds for $Z_b^{\prime}\rightarrow VP$. At the same $\alpha$, the predicted branching ratios of $Z^{(\prime)}_b\rightarrow \rho^+\eta$ are larger than those of $Z_b^{(\prime)}\rightarrow \rho^+\eta^\prime$, which is mainly due to the larger $n\bar n$ component in $\eta$. This point need experimental confirmation in the future. From Eq.~(\ref{eq:eta-etap mixing}), we can obtain that $\theta_P=-19.1^\circ$ corresponds to $66\%$ $n\bar n$ component in $\eta$ and $34\%$ $n\bar n$ component in $\eta^\prime$. In our calculations, we consider $Z_b$ and $Z^{\prime}_{b}$ as $B{\bar B}^*+c.c.$ and $B^*{\bar B}^*$ molecular states, respectively. We only calculate the contribution from neutral and charged bottomed meson loops, thus the larger branching ratio of $Z^{(\prime)}_b\rightarrow \rho^+ \eta$ larger than that of $Z^{(\prime)}_b\rightarrow \rho^+ \eta^{\prime}$ is understandable.
In the range of $\alpha=0.1 \sim 0.3$,  the predicted branching ratios are about 
\begin{eqnarray}
\mathcal{B}[Z_b\rightarrow  \omega\pi^+]&=&(0.58\sim 14.3)\times 10^{-2},\nonumber\\
\mathcal{B}[Z_b\rightarrow \rho^+ \eta]&=&(0.38 \sim 9.40)\times 10^{-2},\nonumber\\
\mathcal{B}[Z_b\rightarrow \rho^+ \eta^\prime]&=&(0.20 \sim 4.80)\times 10^{-2},\nonumber\\
\mathcal{B}[Z_b^{\prime}\rightarrow  \omega\pi^+]&=&(0.66 \sim 16.4)\times 10^{-3},\nonumber\\
\mathcal{B}[Z_b^{\prime}\rightarrow \rho^+ \eta]&=&(0.44 \sim 11.0)\times 10^{-3},\nonumber\\
\mathcal{B}[Z_b^{\prime}\rightarrow \rho^+ \eta^\prime]&=&(0.23 \sim 5.64)\times 10^{-3}. \label{eq:data range}
\end{eqnarray}

It would be interesting to further clarify the uncertainties arising from the introduction of the form factors by studying
the $\alpha$ dependence of the ratios among different branching ratios. For the decays $Z_b \to VP$, we define
the following ratios to the branching ratios of $Z_b \to \omega \pi^+$:
\begin{eqnarray}
	R_{1}&=& \frac{\mathcal{B}[Z_{b}\rightarrow \rho^+ \eta]}{\mathcal{B}[Z_{b}\rightarrow \omega\pi^+]} \, ,\nonumber\\
    R_{2}&=& \frac{\mathcal{B}[Z_{b}\rightarrow \rho^+ \eta^\prime]}{\mathcal{B}[Z_{b}\rightarrow \omega\pi^+]} \, . \label{Eq:ratio-Zb}
\end{eqnarray}
Similarily, for the decays $Z_b^{\prime} \to VP$ the ratios are defined as
\begin{eqnarray}
    r_{1}&=& \frac{\mathcal{B}[Z^{\prime}_{b}\rightarrow \rho^+ \eta]}{\mathcal{B}[Z^{\prime}_{b}\rightarrow \omega\pi^+]} \, ,\nonumber\\
    r_{2}&=& \frac{\mathcal{B}[Z^{\prime}_{b}\rightarrow \rho^+ \eta^\prime]}{\mathcal{B}[Z^{\prime}_{b}\rightarrow \omega\pi^+]} \, . \label{Eq:ratio-Zbp}
\end{eqnarray}

\begin{figure}[t]
  \centering
 \includegraphics[width=8cm]{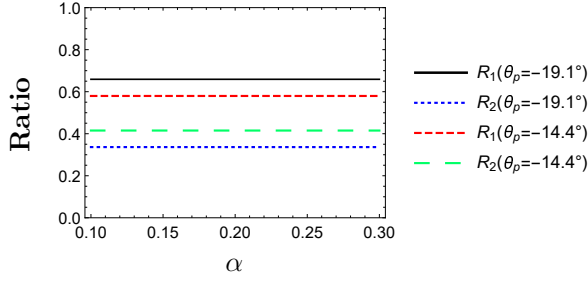}
  \caption{The $\alpha$ dependence of the ratios $R_1$  and $R_2$ defined in Eq.~(\ref{Eq:ratio-Zb}) with $\theta_P=-19.1^\circ$~\cite{MARK-III:1988crp,DM2:1988bfq} and $\theta_P=-14.4^\circ$~\cite{Ambrosino:2009sc}.}\label{Fig:ratio}
\end{figure}

The ratios $R_i$ and $r_i$ $(i=1, 2)$ in terms of $\alpha$ are plotted in Fig.~\ref{Fig:ratio}. On the one hand, it shows that the ratios are almost independent of $\alpha$, which indicates a reasonably controlled cutoff for each channel by the form factor to some extent. On the other hand, one can see that there is a certain dependence of the ratio on the $\eta-\eta^\prime$ mixing angle $\theta_P$, which is of more fundamental significance than the
parameter $\alpha$. This finding stimulates us to study the mixing angle $\theta_P$ dependence.

\begin{figure}[b]
  \centering
 \includegraphics[width=6.5cm]{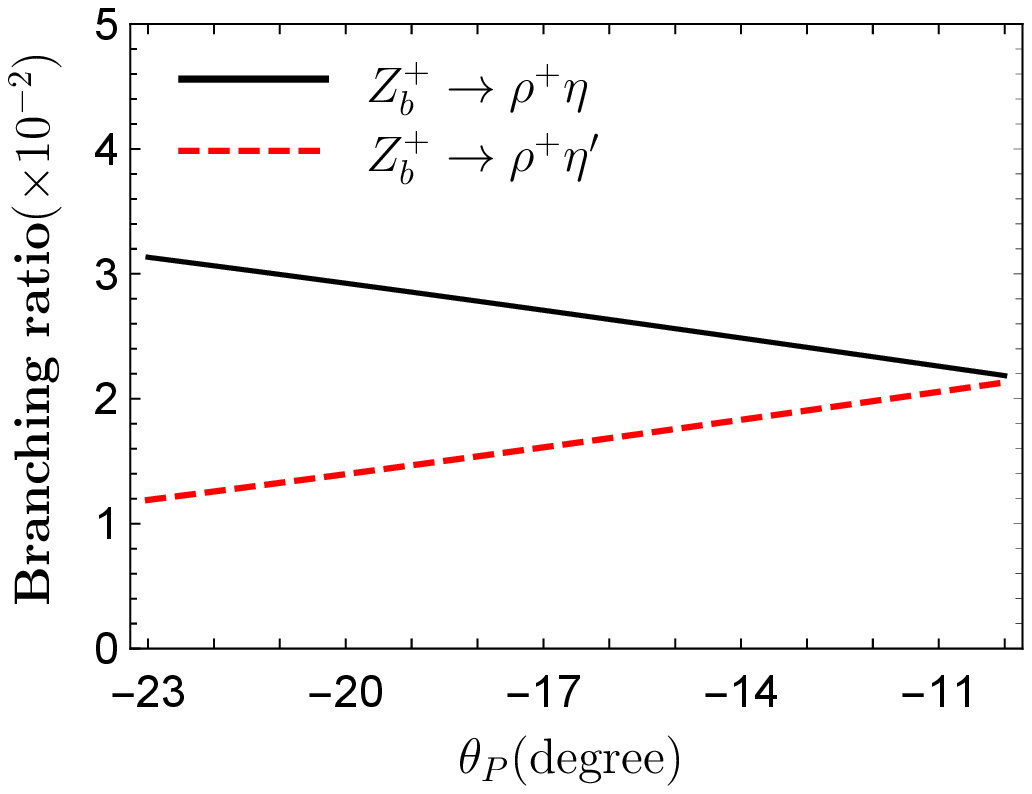}
  \caption{Branching ratios of $Z_b\rightarrow \rho^+ \eta^{(\prime)}$ depending on the mixing angle $\theta_P$ with $\alpha=0.2$.}\label{Fig:B3}
\end{figure}

\begin{figure}[htb!]
  \centering
 \includegraphics[width=6.5cm]{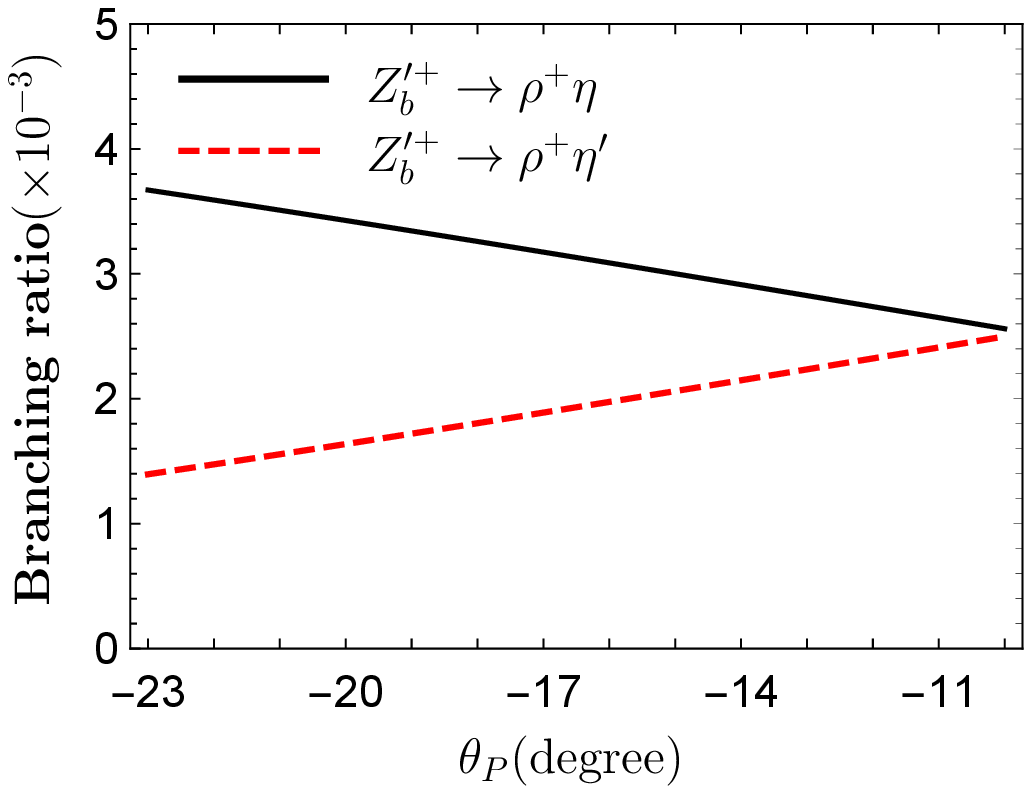}
  \caption{Branching ratios of $Z_b^{\prime}\rightarrow \rho^+ \eta^{(\prime)}$ depending on the mixing angle $\theta_P$ with $\alpha=0.2$.}\label{Fig:B4}
\end{figure}

As is well known, the $\eta-\eta^\prime$ mixing is a long-standing question in history. The mixing angle plays an important role in physical processes involving the $\eta$ and $\eta^\prime$ mesons. However, the mixing angle can neither be calculated from first principles in QCD nor directly measured from experiments. Next, we mainly focus on the impact of the mixing angle on the processes involving the $\eta$ and $\eta^\prime$ mesons, namely $Z^{(\prime)}_{b}\rightarrow \rho^+ \eta^{(\prime)}$. In Figs.~\ref{Fig:B3} and \ref{Fig:B4}, we plot the branching ratios of $Z_b\rightarrow \rho^+ \eta^{(\prime)}$ and $Z^\prime_b\rightarrow \rho^+ \eta^{(\prime)}$ in terms of the $\eta-\eta^\prime$ mixing angle with $\alpha=0.2$. As shown in Fig.~\ref{Fig:B3}, when increasing the mixing angle $\theta_P$, the branching ratio of $Z_b\rightarrow \rho^+ \eta^{\prime}$ increases, while the branching ratio of $Z_b\rightarrow \rho^+ \eta$ decreases. This behavior suggests how the mixing angle influences our calculated results to some extent. As for $Z^{\prime}_b\rightarrow \rho^+ \eta^{(\prime)}$ in Fig.~\ref{Fig:B4}, the situation is similar.

\begin{figure}[b]
  \centering
 \includegraphics[width=8cm]{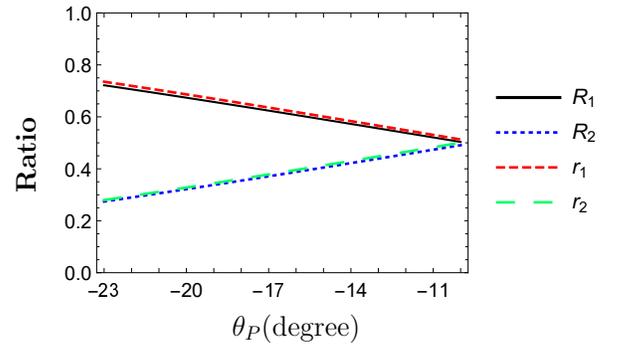}
  \caption{Ratios defined in Eqs.~(\ref{Eq:ratio-Zb}) and  (\ref{Eq:ratio-Zbp}) depending on $\theta_P$ with $\alpha=0.2$.}\label{Fig:ratiomix}
\end{figure}

In Fig.~\ref{Fig:ratiomix}, we present the ratios defined in Eqs.~(\ref{Eq:ratio-Zb}) and (\ref{Eq:ratio-Zbp}) as a function of the mixing angle $\theta_P$ with $\alpha=0.2$. It is interesting to see that the line shape behaviors of the ratios $R_1$ and $r_1$
are almost coincident and the same behaviors for the ratios $R_2$ and $r_2$. The $n\bar n$ components in $\eta$ and $\eta^\prime$ and the intermediate bottomed meson loops may mainly influence these ratios. With the mixing angle range from $-23^\circ$ to $-10^\circ$, the ratios $R_1$ and $r_1$ decrease from $0.73$ to $0.50$, while the ratios $R_2$ and $r_2$ increase from $0.27$ to $0.50$. It is remarkable that 
these four ratios would be approximately equal at $\theta_P \simeq -10^\circ$. We expect the future experiments could measure the ratios in Eqs.~(\ref{Eq:ratio-Zb}) and  (\ref{Eq:ratio-Zbp}), which may help us constrain this mixing angle.

\section{Summary}
\label{sec:summary}

In this work, we evaluated the branching ratios of $Z^{(\prime)}_b \to VP$ by considering the contributions of intermediate bottomed meson triangle loops within an effective Lagrangian approach, where the $Z_b$ and $Z^\prime_b$ are assigned as $B\bar{B}^*+c.c.$ and $B^*\bar{B}^*$ molecular states, respectively. In the calculations, the quantum numbers of these two resonances are fixed to be $I^G(J^P)=1^+(1^+)$, which is consistent with the experimental analysis. The predicted branching ratios of $Z_b\rightarrow VP$ are about in the order of $10^{-2}$, while the branching ratios of $Z^{\prime}_b\rightarrow VP$ are in the order of $10^{-3}$. Moreover, the dependence of these ratios between different decay modes of $Z^{(\prime)}_b$ on the mixing $\eta-\eta^\prime$ angle $\theta_P$ is also investigated, which may be a good quantity for the experiments. It is hoped that the calculations here can be tested by future experiments and can be used to determine the value of the mixing angle.


\begin{acknowledgments}

This work is partly supported by the National Natural Science Foundation of China under Grant Nos.
12075133, 12105153, and 11835015. the Natural Science
Foundation of Shandong Province under Grant No. ZR2021MA082. It is also supported by Taishan
Scholar Project of Shandong Province (Grant No.tsqn202103062),
the Higher Educational Youth Innovation Science and Technology
Program Shandong Province (Grant No. 2020KJJ004).

\end{acknowledgments}

\end{document}